\documentclass[nofootinbib,12pt]{revtex4-1}
\usepackage{amsmath,amssymb,pgf,pgfarrows,pgfnodes,float,appendix, hyperref}
\usepackage{graphicx}
\usepackage{subfigure}
\usepackage[margin=1in]{geometry}

\newcommand{\TR}{\mathop{\rm Tr}}

\newcommand{\calO}{{\cal O}}

\newcommand{\bea}{\begin{eqnarray}}
\newcommand{\eea}{\end{eqnarray}}

\begin{document}
\title{Comments on the Entanglement Spectrum of de Sitter Space}
\author{Tom Banks}
\affiliation{NHETC and Dept. of Physics and Astronomy, Rutgers University, Piscataway, NJ }
\author{Patrick Draper}
\affiliation{Department of Physics, University of Illinois, Urbana, IL 61801}

\begin{abstract}
We argue that the Schwarzschild-de Sitter black hole entropy formula does not imply that the entanglement spectrum of the vacuum density matrix of de Sitter space is flat. 
Specifically, we show that the expectation value of a random projection operator of dimension $d\gg 1$, on a Hilbert space of dimension $D\gg d$ and in a density matrix $\rho = e^{-K}$ with strictly positive spectrum, is $\frac{d}{D}\left(1 + o(\frac{1}{\sqrt{d}})\right)$, independent of the spectrum of the density matrix.  In addition, for a suitable class of spectra the asymptotic estimates ${\rm Tr} (\rho K) \sim {\rm ln}\ D - o(1)$ and $ {\rm Tr} [\rho (K - \langle K\rangle)^2] = a \langle K \rangle$ are compatible for any  order one constant $a$.   We discuss a simple family of matrix models and projections that can replicate such modular Hamiltonians and the SdS entropy formula.

\end{abstract}

\maketitle

\section{Introduction}
The Covariant Entropy Bound (CEB)~\cite{fsb} is the conjecture that a causal diamond in an arbitrary space-time  is described by a quantum system with maximal entropy $\frac{A}{4G}$, where $A$ is the maximal $d - 2$ volume of a space-like leaf in a null foliation of the diamond's boundary.  Bousso~\cite{raph} conjectured that the  equilibrium density matrix is maximally uncertain in the finite dimensional Hilbert space of the diamond.  Fischler~\cite{wfdS} and one of the present authors~\cite{tbdS} applied this conjecture to de Sitter space.  It was also noted in~\cite{wfdS}  that the Schwarzschild de Sitter black hole entropy formula implies that maximal entropy corresponds to empty de Sitter space, so that localized objects in a causal patch of de Sitter space correspond to constrained, low entropy states. In terms of the gravitational path integral, this means that Euclidean SdS is a semiclassical solution subject to a constraint on the quasilocal energy~\cite{Draper:2022xzl}\cite{Morvan:2022ybp}. The migration of a typical localized object to the cosmological horizon, where it becomes absorbed in the featureless background, is then understood as equilibration of this low entropy state.

Some years later, a class of models of finite quantum systems were shown to reproduce the SdS entropy formula in four dimensions~\cite{bfm}. The relevant systems are matrix models of fermionic oscillators with single-trace Hamiltonians.  The constrained subspace corresponding to a black hole is one in which certain sets of fermionic oscillators vanish.  Fischler and Banks used similar constructions to sketch a theory of scattering in Minkowski space~\cite{tbwfscatt}, and a novel model of inflationary cosmology~\cite{tbwfinflation}.

The conjecture that the equilibrium density matrix is maximally uncertain  was motivated by the supposed generality of the CEB. General space-times do not have time translation symmetry, and causal diamonds are not invariant under time translation, so no natural thermal density matrix is available.  The argument was essentially "What else could it be?"\footnote{Fischler and TB later came up with what we thought was a better argument, based on Jacobson's use of an infinitely accelerated Unruh trajectory\cite{ted95} to derive Einstein's equations from (essentially) the CEB and the local first law of thermodynamics.   This argument is incorrect.  The modular Hamiltonian of a diamond is independent of the trajectory.  The differences between different trajectories all have to do with relative time dilation.}. 

More recently, work on the AdS/CFT correspondence has led to a better answer to the question of what the density matrix could be.  In an important paper, Casini, Huerta and Myers~\cite{CHM} showed that the modular Hamiltonian of a Minkowski diamond in a conformal field theory (CFT) is the quantum generator of the action of a conformal Killing vector that preserves the diamond, evaluated on the holographic screen of the diamond.  Using the AdS/CFT correspondence, they showed that this is the same as the diffeomorphism that preserves the Ryu-Takayanagi (RT) diamond anchored to the Minkowski diamond.  Zurek and TB~\cite{BZ} conjectured a similar statement for non-extremal diamond boundaries in arbitrary space-times, when the quantum gravity model is well approximated by the Einstein-Hilbert action.  Ref.~\cite{BZ} generalized seminal work by Carlip~\cite{carlip} and Solodukhin~\cite{solo}, who argued that the modular Hamiltonian of quite general black holes is the $L_0$ generator of a Virasoro algebra.  This leads to the entropy fluctuation formula
\begin{equation} \langle (K - \langle K \rangle)^2 \rangle = \langle K \rangle . \end{equation}  That formula was previously established~\cite{VZ2} for RT diamonds in arbitrary CFTs with an Einstein-Hilbert dual.  

Several recent papers~\cite{evaetal}\cite{Chandrasekaran:2022cip}\cite{susshyper} have revived the conjecture that the entanglement spectrum of dS space is flat.  They appear to have all been motivated in part by the SdS black hole entropy formula.  The purpose of the present  note is to show that that motivation is insufficient.  Abstractly, the key message of \cite{bfm} was that for a flat entanglement spectrum, the quantum probability of being in a $d$-dimensional subspace of a $D$-dimensional Hilbert space is $\frac{d}{D}$, to leading order when $1\ll d \ll D.$\footnote{Note that the classical probability is zero.  In quantum mechanics one only requires that the state has an $o(1)$ projection on the subspace.}  We will show  that the same result holds, up to corrections of relative order $\frac{1}{\sqrt{d}}$, for any density matrix whose modular Hamiltonian has a finite spectrum, and for a random choice of the projector.    Of course, if the entanglement spectrum is not flat, there will be {\it some} projectors for which this estimate is incorrect.  Our result shows that they are atypical.

One might want to retain another characteristic of the vacuum dS density matrix, namely that
\begin{equation} \langle K \rangle = {\rm ln}\ D + \calO(1) \label{logD1}.\end{equation}  We will describe a simple class of density matrices for which this property holds and that are compatible with any fluctuation formula of the form
\begin{equation} \langle (K - \langle K \rangle)^2 \rangle = a \langle K \rangle , \end{equation}  for $a$ of order $1$.

We note that insisting on an order-one correction to $S={\rm ln}\ D$ as in Eq.~(\ref{logD1}) may be too strong a requirement.  There is a long history of computations in the literature that suggest the  black hole entropy formula has corrections of order ${\rm ln}\ S$. Similarly in Jacobson's derivation of Einstein's equations from the first law one expects subleading corrections to the relation between entropy and area due to higher derivative terms in the gravitational action.  We will see that having an order-one correction to $S = {\rm ln}\ D$ in our models requires a fine-tuning of the UV cutoff on the entanglement spectrum so that it is close to a saddle point in the coarse-grained partition function.   We could modify our results to obtain ${\rm ln\, ln\,} D$ corrections by reducing this fine tuning.  A principled motivation for the tuning must be the subject of further work.

\section{Quantum Probability of Random Projections}
Let  $\rho_i$ be the eigenvalues of the density matrix on a Hilbert space of dimension $D$. Then define the random $d$-dimensional projection
\begin{align}
P_d = \sum_{a=1}^d U_{ia} U_{aj}^\dagger
\end{align}
where $U$ is a random unitary. Let $\langle\langle\rangle\rangle$ denote averaging over unitaries. Then the probability to be in a state in the projection subspace is, on average,
\begin{align}
p=\langle\langle \TR (\rho P_d)\rangle\rangle = \sum_{i=1}^D\sum_{j=1}^d\rho_i \langle\langle U_{ij}U^\dagger_{ji}\rangle\rangle = \frac{d}{D}\sum_{i=1}^D\rho_i = \frac{d}{D}.
\end{align}
The fluctuation in this probability is 
\begin{align}
\sigma^2\equiv \langle\langle \TR (\rho P_d) \TR (\rho P_d)\rangle\rangle - \langle\langle  \TR (\rho P_d)\rangle\rangle^2 = \sum_{i,j=1}^D\sum_{k,l=1}^d \rho_i\rho_j \langle\langle U_{ik}U^\dagger_{ki}U_{jl}U^\dagger_{lj}\rangle\rangle - \left(\frac{d}{D}\right)^2.
\end{align}
The term involving the average over four unitaries is
\begin{align}
\sum_{k,l=1}^d\langle\langle U_{ik}U^\dagger_{ki}U_{jl}U^\dagger_{lj}\rangle\rangle = d^2 W_D(1^2) + d W_D(2) +\delta_{ij}(d^2 W_D(2) + d W_D(1^2)).
\end{align}
Here $W_D(\tau)$ is the Weingarten function \cite{weingarten} associated with a permutation of the indices of the unitaries with cycle shape $\tau$. For four unitaries the relevant permutation group is just $S_2$, and
\begin{align}
W_D(1^2)& = \frac{1}{D^2-1}\nonumber\\
W_D(2) &= -\frac{1}{D^3-D}.
\end{align}
All together we have
\begin{align}
\sigma^2 = \left( \frac{d^2}{D^2-1}- \frac{d}{D^3-D} -\frac{d^2}{D^2}\right) + \left(\frac{d}{D^2-1}-\frac{d^2}{D^3-D}\right)\sum_i \rho_i^2 .
\end{align}
For $D\gg d\gg 1$,
\begin{align}
\sigma^2 =\frac{d}{D^3} + \frac{d}{D^2} \sum_i \rho_i^2 .
\end{align}
For a pure state, $\sum_i \rho_i^2=1$, and
\begin{align}
\sigma^2_{\rm \tiny pure} = \frac{d}{D^2}+\calO\left(\frac{d^2}{D^3}\right).
\end{align}
The relative fluctuation in this case is $\sqrt{\sigma^2}/p\sim 1/\sqrt{d}$. 
For a maximally uncertain state, $\sum_i \rho_i^2=1/D$, and
\begin{align}
\sigma^2_{\rm \tiny max} = 0.
\end{align}
We can conclude that for typical projections and general states, the probability of being in the subspace compatible with the projection is very close to $d/D$, for $1\ll d\ll D$.

\section{Modular Fluctuations}
Let $\rho = e^{-K}$. The eigenvalues are $\rho_i = e^{-k_i}$. The entropy  is
\begin{align}
S= -\TR(\rho\log\rho) = \sum_i k_i e^{-k_i}  =\langle K\rangle.
\end{align}
We would like this to be of order $\log D$, with $\sum_i e^{-k_i}=1$. A simple solution is to be near the maximally uncertain state. Write $k_i= \log D-\delta_i$ with $\delta_i\sim\calO(1)$. Then the modular expectation value and fluctuations are
\begin{align}
&\langle K\rangle= \log D+\calO(1)\;,\nonumber\\
&\langle(\Delta K)^2\rangle=\langle(\Delta \delta)^2\rangle\sim \calO(1).
\end{align}
One way  to have larger fluctuations is to relax the requirement on $\langle K\rangle$, allowing $\langle K\rangle \sim \log D$, instead of $\langle K\rangle =\log D + ({\rm \tiny small})$. For example, take a simple system of $N$ independent qubits with gap $\Delta E$ in a thermal state of temperature $T$. The maximum entropy is $\log D = N\log 2$, achieved for $\Delta E/T\rightarrow 0$. For any $\Delta E/T\sim\calO(1)$, the entropy and the modular fluctuation are both $\calO(N)$.

But there are other interesting possibilities. Now consider density matrices with spectra that have level splittings that decrease rapidly as the modular eigenvalue $k_i$ increases.  Define the average spectral density at some point $k$ on the positive real line to be 
\begin{equation}\sigma (k) = \frac{1}{2\delta}\int_{k - \delta}^{k + \delta}dk' \sum_i \delta (k' - k_i)  \end{equation}
for some suitable small $\delta$, and consider states for which
\begin{align} \sigma (k) \approx \theta (k - k_1) \theta (k_2 - k) e^{a k^p }, ~~~~~~0 < p < 1,~~~~a \gg 1.
 \end{align} 
For $p = \frac{d - 1}{d} $ this is the behavior of the density of states of a $d$-dimensional quantum field theory at asymptotic energies.\footnote{The requirement $a \gg 1$ is the analog of ``large central charge" in arbitrary dimension.  In the context of the AdS/CFT correspondence $a \gg 1$ is a necessary condition for even a ``stringy" interpretation of the bulk space-time.}

For such  spectral densities, $\log\,D\approx a k_2^p$. The entropy and its fluctuations can be evaluated in the saddle point approximation. Write the normalized averaged density matrix as $\rho(k) = N e^{-k}$. Then
   \begin{align}
   1 = \int dk\, \sigma(k) \rho(k) = N\int_{k_1}^{k_2} dk \,e^{a k^p -k}.
   \end{align}
  The saddle point is $k_\star= (ap)^{\frac{1}{ 1-p}} $, so $N\approx e^{k_\star-a k_\star^p}$ and
  \begin{align}
  \rho = e^{-(k-k_\star)-a k_\star^p} \equiv e^{-K(k)}.
  \end{align}
  The entropy is
  \begin{align}
  S = -\TR(\rho\log\rho) =   \langle K \rangle = a k_\star^p =  \frac{1}{p}(ap)^{\frac{1}{ 1-p}} .
  \end{align}
  The modular fluctuations are easily computed by introducing a fictitious temperature,
  \begin{align}
  Z(\beta) = \int dk \,\sigma(k) e^{-\beta K(k)}.
  \end{align}
  Now the saddle point is  $k_\star= (ap/\beta)^{\frac{1}{ 1-p}} $, so 
   \begin{align}
   \log Z(\beta) \approx a \left(\frac{ a p}{\beta}\right)^\frac{p}{1-p} - \beta  \left(\frac{ a p}{\beta}\right)^\frac{1}{1-p} +({\rm linear~in~}\beta).
   \end{align}
We have
  \begin{align}
  \langle (\Delta K)^2\rangle = \partial_\beta^2 \log Z\big|_{\beta=1} = \frac{(ap)^{\frac{1}{ 1-p}} }{1-p}.
   \end{align}
   The ratio is 
   \begin{align}
   \frac{\langle (\Delta K)^2\rangle}{ \langle K \rangle} = \frac{p}{1-p}.
   \end{align} 
   Thus we see it is possible to obtain both an entropy close to $\log D$ {\emph{and}} large entropy fluctuations, i.e. in a  state where the entanglement spectrum is not flat. 
In particular for the case of a 1+1D CFT, $p=1/2$, we recover the result $ (\Delta K)^2  =S $ of~\cite{BZ}.

 For general density matrices of this form, imposing  $S = \log D - \calO(1)$ requires a fine tuning of the cutoff $k_2$ close to the saddle $k_*$. On the other hand, if one of the axioms of the eventual general theory of quantum gravity is that the modular Hamiltonian of a diamond is a cutoff Virasoro generator, it is simply the statement that no larger Hilbert space is necessary to explain the physics of horizons.

\section{More Realistic Models}  In \cite{bfm} models, of four dimensional dS space were proposed in which the variables are $R \times R+1$ matrices, $\psi_i^A$, the matrix elements of which are independent fermionic oscillators.  $R$ is an integer proportional to the dS radius in Planck units.  The Hamiltonian for propagation between a causal diamond with entropy $\pi\rho^2 $ and one with entropy $\pi (\rho + 1)^2$ has the form
\begin{equation} H = \rho^{-1} {\rm Tr}\ P(M/\rho) , \end{equation} where $M_i^j = \psi_i^A \psi^{\dagger\ j}_A $, is constructed from a $\rho \times (\rho + 1)$ subset of the variables.  $\rho \rightarrow R$ at infinite proper time. 
It was proposed that the vacuum density matrix is maximally uncertain and that states of objects localized in the static patch are states in which of order $ER$ of the oscillators are set to zero, with $E$ proportional to the non-conserved static energy. The energy is not conserved because all of these localized states eventually relax to the vacuum equilibrium, where most of the states have eigenvalues of the static Hamiltonian that are of order $1/R$.  In the limit $R\rightarrow\infty$, $E$ becomes the asymptotically conserved energy of flat space-times. The form of the (time dependent) Hamiltonian guarantees that $E$ is approximately conserved on time scales at least as long as $R$.  

Insights based on the HST model of inflationary cosmology \cite{holoinflation1+1}, and the work of Carlip and Solodukhin\cite{carlip}\cite{solo}\cite{BZ}, suggest instead that the modular Hamiltonian is the $L_0$ generator of a Virasoro algebra, in a unitary lowest weight representation with large central charge and a UV cutoff.
We can modify the setup of \cite{bfm} to incorporate this possibility by allowing the fermions to depend on another index $\psi_i^A (m)$.  $m$ runs from $1$ to $M$, where $1 \ll M \ll R$.  We write a quadratic formula for the {\it modular} Hamiltonian, of the form
\begin{equation} K = \psi^{\dagger\ i}_A (m) h_{mn} \psi_i^A (n) , \end{equation} where $h$ is a random $M\times M$ Hermitian matrix.  It is well-known~\cite{kapwein}, as a consequence of Wigner's semi-circle law,  that if $h$ is chosen from an ensemble close to Gaussian, this system is, in the large $M$ limit, equivalent to the CFT of $R(R + 1)$ free fermion fields on an interval, with a UV cutoff determined by the form of the distribution for $h$.  

Our subspace is defined by
\begin{equation} {M}_i^j (n) | s\rangle = 0, \end{equation} for all $n$, $i = 1 \ldots E$, and $j = E + 1 \ldots R $, as well as the (matrix) Hermitian conjugate of this equation.   Diagonalizing the matrix $h$, the subspace density matrix is
\begin{equation} \otimes^{ER} \frac{ \prod_{m=1}^M e^{- h (m) n(m)}}{\prod_{m=1}^M (1 + e^{-h(m)})} \end{equation} 
 and the projector is onto  the state with all $n(m) = 0$.  In this formula $n(m)$ in the $iA$ factor in the product is the number operator $\psi^{\dagger A}_i \psi^i_A$, with no summation over indices.
The probability of this state in the density matrix $e^{-K}$  is
\begin{equation} P = \left(\prod_{n=1}^M (1 + e^{-h(m)})\right)^{-ER} , \end{equation} 
which certainly has a Boltzmann form with temperature proportional to $R^{-1}$ .  So for this simple model, we reproduce the Schwarzschild de Sitter entropy formula by insisting that $E\, {\rm ln} \prod_{n=1}^M (1 + e^{-h(m)})$ is the energy in Planck units, when  $E \ll R$.   The total entropy satisfies the criterion of the previous section (with $p = 1/2$) when $h$ is taken to be a random $M \times M$ matrix, with $M \gg 1$, if the UV cutoff in the distribution from which $h$ is chosen is tuned so that the entropy is $ S = MR(R + 1) + \calO(1)$ as $R \rightarrow \infty$.    
There are a variety of exactly marginal perturbations of the free fermions, which should preserve all of these properties for sufficiently weak coupling. 

\section{Discussion}

The results of the previous section show that the fact that localized states are low entropy constrained subspaces of the Hilbert space of dS space does not by itself tell us much about the dS vacuum density matrix.  In fact, the proposals of \cite{bfm} had many more features that reproduced semi-classical properties of dS.  The Hamiltonian was assumed to be a single trace of a polynomial in matrices that were built from fermion bilinears,
$M_{i}^j = \psi_i^A \psi^{\dagger\ j}_A $.
The constrained states were defined by insisting that the Hamiltonian acting on them was the trace of a block diagonal matrix, with a number of small blocks of size $E_i$ and one much larger block of size $R/L_P - \sum E_i$.   As a consequence the blocks behave for some time like independent systems.  Moreover, the residual $S_N$ gauge invariance on $N$ identical blocks can be identified with the statistical gauge invariance of quantum particles.  Finally,  single trace matrix Hamiltonians are almost certainly fast scramblers~\cite{hpss}, and this is as much a property of the dS horizon as it is of small black hole horizons. 

Another reason to question the flatness of the dS entanglement spectrum is that, with the exception of~\cite{evaetal}, the arguments for it apply equally well to arbitrary finite area causal diamonds.  For finite diamonds, this is in conflict with the elegant results of \cite{CHM}, which when combined with the results of \cite{carlip} and \cite{solo} provide a universal\footnote{Up to the choice of a $1+1$ dimensional CFT with large central charge.} \cite{BZ} prescription for the modular Hamiltonian of a causal diamond. 

All of the new arguments for flatness of the entanglement spectrum are leading order calculations in a large entropy expansion.  The paper of~\cite{Chandrasekaran:2022cip} does not work within a framework where finite $S$ corrections are calculable.   In the other two approaches, the leading finite $S$ corrections have not yet been calculated.  If they disagree with the answer suggested in~\cite{BZ}, we will have several different models of dS space, all of which account for the semiclassical facts about its entropy and the entropy of its excitations.

\vspace{1mm}

\begin{center}
{\bf Acknowledgments }\\
\end{center}

The work of TB is partially supported by the Department of Energy under grant DOE SC0010008. PD acknowledges support from the US Department of Energy under Grant number DESC0015655.

\end{document}